\documentclass[nohyper,12pt,letterpaper]{JHEP}
\usepackage{epsfig}

\DeclareSymbolFont{AMSa}{U}{msa}{m}{n}
\DeclareSymbolFont{AMSb}{U}{msb}{m}{n}
\let\Box\relax
\DeclareMathSymbol{\Box}{\mathord}{AMSa}{"03}

\newcommand{\vdot}{\dot{v}}
\newcommand{\Zdot}{\dot{Z}}

\def \eqn#1#2{\begin{equation}#2\label{#1}\end{equation}}

\title{The Entropy of the Microwave Background and the Acceleration
of the Universe}

\author{W. Fischler, A. Loewy, S. Paban\\
Department of Physics \\ University of Texas, Austin,
TX 78712}

\abstract{If the present acceleration of the universe is due to a
cosmological constant, $\lambda$, then the entropy of the microwave
   background is bounded. It cannot exceed
$\lambda^{-3/4}\sim 10^{91}$, which is much less than the entropy of
empty de Sitter
space $\lambda^{-1}\sim10^{122}$.  This is due
to the limited efficiency of storing entropy by local field
theoretical degrees of
freedom. The observed entropy of the microwave background is of
$O(10^{85})$.}


\received{???????? ?st, 2000} \accepted{???????? ?th, 1998}
\preprint{\hepth{0307031}\\UTTG-03-03}

\begin{document}




\section{\bf Introduction}

Recent observations of supernovae indicate that the universe is
accelerating \cite{Perlmutter:1997ds,Riess:1998cb,Spergel:2003cb}. 
This suggests that, at present, the energy
density of the universe is dominated by
a fluid with an equation of state $p < - \frac{1}{3} \; \rho$.
We will assume in this paper that the source of the observed
acceleration is a positive
cosmological constant
\eqn{lstate}{p = -\rho = -\lambda \ .}
We will then argue that in this case, the entropy stored in local field
theoretical degrees of freedom cannot exceed a bound set by a power
of the cosmological
constant. The upper bound on this entropy scales like ${\lambda}^{-3/4}$, which
is parametrically smaller than the entropy of empty de Sitter space,
$S_{DS}\sim
{\lambda}^{-1}$ \cite{GH}.
We will refer in the rest of the paper to the entropy of the
microwave background,
although the discussion applies to the entropy of any field
theoretical degree of freedom.

The physics that underlies the bound, in a nutshell, is the limited
efficiency with which a
fluid described by local field theoretical degrees of freedom stores
entropy in space. This
capacity to store information, as seen by local observers, depends on
the equation of state.
It was shown
\cite{zurekpagegerardstars} that the entropy of a fluid with an equation of state
$ p =\kappa\rho$ scales as a power of the linear size $R$,
\eqn{R^}{ S \sim  R^{\ 3 - \frac{2}{1 + \kappa}} \ .}
This result was obtained in the absence of a cosmological constant.
We will show that any attempt, when there is
no cosmological constant, to squeeze more entropy in a region of
fixed size $R$ leads to the formation of black
holes.
This implies that for radiation, $\kappa = 1/3$, a given region of
space with linear size $R$ can only
accommodate an entropy of $O(R^{3/2})$. This is rather inefficient
when compared to the storage
capabilities of a black hole, or a fluid with $\kappa = 1$. Indeed,
in the latter cases the entropy scales like
the area, $R^2$.

One might then wonder about what happens to the ``storage of entropy''
in the presence of a positive
cosmological constant. We will argue that in order for such a
space-time 
to asymptote to de Sitter space,
there is a bound on the amount of stored entropy, that a local
observer can ever measure. This local
observer eventually perceives this space-time as a hot cavity
\cite{lennybf} of finite size ${\lambda}^{-1/2}$ in
which the entropy of the microwave background never exceeds
${\lambda}^{-3/4}$. The bulk of the paper is devoted
to argue that any attempt to store more entropy than the bound
allows, meets with black hole formation or a
change of the equation of state to a stiffer one, or more probably
imprints itself on the event horizon of de
Sitter space, and in some cases the universe ends up in a big crunch.

The paper is organized as follows:
In section 2, as a warm-up, we discuss what happens in the contracting phase
of a space-time with a cosmological constant and a perfect fluid, as a
function of the entropy of the fluid.
In section 3, we discuss how efficient a fluid in equilibrium can
be in packing entropy.  We then use in section 4 the
Tolman-Oppenheimer-Snyder (TOS) model \cite{TOSeverybody} in a way
that differs from its original purpose.
In this paper, we use  the TOS model in the presence of a
cosmological constant \cite{TOSlambda}
to give us a handle on the amount of entropy inside a ``star''. This
will allow us to draw further evidence for our bound.
We end with conclusions and an appendix where a preliminary setup in
the ``O-gauge'' \cite{LennyO} can be found.


\section{\bf Contracting Phase of a Space-time with a Fluid and a
Cosmological Constant}

In this section we will consider the case of a homogeneous
universe with a positive cosmological constant and a perfect fluid.
We will show that any attempt to store more
information, during the contracting phase of this cosmology, than what
is allowed by a bound on the entropy,
leads to a big crunch \footnote{For a discussion on the endpoint of a
big crunch, see
\cite{blackcrunch}.}.
Consider he Friedmann-Robertson-Walker (FRW) metric,
\eqn{metric}{ds^2 = dt^2 - a(t)^2 \ \left(d\theta^2 +
\sin^2{\theta}d\Omega^2_2 \right) \ .}
The equation that the scale factor obeys is,
\eqn{a(t)}{ \left(\frac{{\dot{a}}}{{a}} \right)^2 + \frac{1} {a^2} =
\frac{\lambda}{3} + \rho(a) \ ,}
where $\rho(a)$ is the energy density of the fluid {\footnote{For the
remainder of the
paper we will work in Planck units.}}.
We will use a fluid  with a linear equation of state, $p = \kappa\rho$.
For a given $\kappa$ the energy density obeys
\eqn{rho(a)}{ \rho(a) = \frac{ \rho(a_0) \ {a_0}^{3(1 + \kappa)}}{a
^{3(1 + \kappa)}}\ .}
By using the thermodynamic relation between the energy and
entropy densities,
\eqn{thermo}{\rho \sim {\sigma}^{1 + \kappa} \ ,}
it is then convenient for our purposes to rewrite the energy density as,
\eqn{rho(S)}{\rho(a) = \frac{{S_0}^{1+\kappa}}{a ^{3(1 + \kappa)}} \ ,}
where  $S_0$ is the entropy associated with the fluid when the scale 
factor is $a_0$.
Using this last equation, we can then rewrite (\ref{a(t)}) as
\eqn{a(S)}{\left(\frac{{\dot{a}}}{{a}}\right)^2 + \frac{1} {a^2} = \frac{\lambda}{3} +
\frac{{S_0}^{1+\kappa}}{a ^{3(1 + \kappa)}} \ .}
This equation  implies that in order to avoid a big
crunch, $a \rightarrow 0$, $S_0$ should be bounded from above by
\eqn{boundS_0}{S_0 < C_{\kappa} {\lambda}^{-\frac{1 + 3\kappa}{2(1 +
\kappa)}} \ , }
where $C_\kappa$ is a positive number that depends on $\kappa$ and
microscopic physics of the fluid.\footnote{A similar bound on the
entropy as a function of the cosmological constant for closed
universes was found years ago by Gibbons \cite{Gibbons} in a
completely different context.}
For radiation the bound reads
\eqn{boundS_photons}{S_0 < C_{1/3} {\lambda}^{-{3/4}} \ ,}
where $C_{1/3}$ is a coefficient of $O(1)$.

The bound for the entropy given by (\ref{boundS_0}), has an
intriguing interpretation. The power law
is exactly the one that is obtained if one answers the question about
the maximum amount of entropy in the fluid,
at the threshold of black hole formation, that can fit in a box of
linear size, ${\lambda}^{-1/2}$, i.e.
the size of the de Sitter horizon.

\section{\bf Efficiency of Entropy Packing}

In this section we will give a heuristic derivation of the
scaling law of the entropy of a fluid as a
function of the size of the region to which it is confined.
We will assume that a perfect fluid contained in a box of size R
has a linear equation of state $p =
\kappa\rho$. We will estimate the entropy versus box size as
one tries to pack entropy to the verge
of black hole formation.  We will assume that a single mode but no
more than a single mode of the fluid,
with energy $\epsilon$ occupies a volume ${\epsilon}^3$ in Planck
units.  The number density of modes is
\eqn{numberd}{ n(\epsilon) = {\epsilon}^{-3} \ .}
The energy density as a function of $\epsilon$ is,
\eqn{energyd}{ \rho(\epsilon) = {\epsilon}^{-2} \ .}
Using the thermodynamics relation between the entropy density $\sigma$
and the energy density $\rho$,
\eqn{entropyd}{\sigma(\epsilon) \sim
{\rho(\epsilon)}^{\frac{1}{1+\kappa}} \ ,}
we obtain a relation, at the threshold of black hole formation
\cite{tom}, between the entropy $S$ and the size $R$
of the region where the fluid is contained.
\eqn{entropyR}{ S(R) \sim \int_0^R d{\epsilon} \ \epsilon^2 \sigma(\epsilon)
  \sim R^{ \ 3 - \frac{2}{1 + \kappa}} \ .}

Notice that this is the exact same scaling law for the entropy that
was derived from the
Tolman-Oppenheimer-Volkov (TOV)  equations
\cite{zurekpagegerardstars} describing a spherically symmetric
perfect fluid in equilibrium \footnote{The derivation of the entropy
at the threshold of black hole formation can be easily
generalized to an arbitrary number of dimensions and $\frac{1}{1-D} <\kappa \le
1$. The
result reproduces the scaling obtained using the TOV
equations in D dimensions}.

It seems reasonable to conclude, based on the earlier heuristic
derivation of the scaling law,  that any
attempt to squeeze more entropy in a given volume beyond the bound
(\ref{boundS_0}) leads to black hole
formation. This picture seems consistent with our present
understanding that a black hole is the most efficient
environment at packing entropy.

The previous discussion on efficient entropy packing, is modified by
the presence of a cosmological constant. Additional distinction should
be made between the contracting and expanding parts of the FRW geometry.
In the contracting
phase black hole formation is enhanced, and the bound on the entropy
remains the same.
In the expanding case, when the size of the
region containing the fluid becomes of
$O({\lambda}^{-1/2})$ the fluid is subjected to a pull due to the
cosmological constant, which makes it
increasingly harder to increase the entropy in that region beyond the
bound discussed in the absence of
$\lambda$. However any attempt to violate the bound is accompanied
by a repackaging of the excess entropy until one
reaches the ``Bekenstein bound''.

In order to provide additional intuition about the bound for the
entropy of a fluid in the presence of a
cosmological constant, we will devote our next discussion to studying
a modified version of the TOS model
\cite{TOSeverybody}.  The TOS model was originally designed to
describe the collapse of a spherical star made of
dust. The interior of the star is modeled by a FRW geometry, whereas
the exterior (using Birkhoff's theorem) is
a static Schwarzchild geometry.

\section{A generalized TOS model}

The TOS model will be modified in what follows, by adding a
cosmological constant
\cite{TOSlambda}. Our intention is to view the star from the outside
as the amount of entropy on
the inside is tuned to eventually violate the aforementioned bound.
In the setup of this generalized TOS
model, we will closely follow some of the analysis by Markovic and
Shapiro \cite{TOSlambda}. This study will give
us evidence for our claim that the entropy in a universe with a
cosmological constant is bounded by an amount
that is parametrically smaller than the Bekenstein bound.

Basically, the TOS model describes the dynamics of a ball of dust in 
the presence of a
positive cosmological constant. The region inside the ball is described
by a (truncated) FRW geometry,
\eqn{intds}{ds^2 = {d\tau}^2 -a(\tau)^2 \left(\frac{dx^2}{1 - kx^2}
+x^2{d\Omega_2}^2 \right) \ , }
where $k$ determines the curvature of the spatial part of the
geometry. The $x$ coordinate take values in the region $0 \le x \le
X$.

The scale factor $a(\tau)$ satisfies
\eqn{hubble}{\left(\frac{{\dot{a}}}{{a}}\right)^2 + \frac{k} {a^2} = \frac{\lambda}{3} +
\frac{\xi}{F(X)} \frac{S(a_0)}{a^3} \ ,}
where $\xi$ is a fixed parameter that has dimensions of mass, and
depends on microscopic physics. This
parameter appears in the relation between the energy density
and the entropy density of the dust
\eqn{dust}{\rho = \xi\sigma \ ,}
and
\eqn{F(X)}{F(X) = \int_0^X \frac{dx \ x^2}{\sqrt{1 - kx^2}}\ .}
The generalized Birkhoff theorem \cite{5ofshapiro} guarantees that
the space-time outside the sphere is
Schwarzchild-de Sitter (SdS)
\eqn{SdS}{ds^2 = \left( 1 - \frac{2M}{r} - \frac{\lambda}{3}r^2 \right) dt^2 -
\frac{dr^2}{\left(1 - \frac{2M}{r} -
\frac{\lambda}{3}r^2 \right)} - r^2 {d\Omega_2}^2 \ .}
In order to complete the description of the space-time geometry, we
need to match the 3-geometry across the
interface between the dust and the exterior space-time, which is at $r
= R(\tau)$ using the the exterior coordinate patch, and at
$x = X$ using the interior coordinate patch. This is
achieved by imposing the continuity of the surface's 3-metric
\eqn{3metric}{ds_{3}^2 = {d\tau}^2 - R^2{d\Omega_2}^2 \ ,}
where $R(\tau) = Xa(\tau)$, and
then imposing Israel's junction conditions \cite{israel}. As shown in
\cite{TOSlambda} these conditions lead to an equation for $R(\tau)$
\eqn{R(tau)}{\dot{R}^2 + f(R) = 1 - kX^2 \ ,}
where
\eqn{f(R)}{f(R) = 1 - \frac{2M}{R} - \frac{\lambda R^2}{ 3} \ .}

We want to remind the reader that we are exploring the consequences
of injecting more entropy than the bound
given in (\ref{boundS_0}). With that in mind we will take an
$S_0$ in (\ref{hubble}) that
exceeds the bound by requiring $S_0 > {\lambda}^{-1/2}$ parametrically.

By comparing equations (\ref{R(tau)}), (\ref{f(R)}) and 
(\ref{hubble}) one finds a relation between
the ``mass'' as seen from the outside, and the amount of entropy,
inside the ``star''
\eqn{M}{ M = \xi \frac{X^3}{F(X)} S(a_0) \ ,}
where the ratio $\frac{X^3}{F(X)}$ is bounded below by a number of
$O(1)$, independent of $\lambda$.

If we now impose that the entropy exceeds our bound, $S_0 >
{\lambda}^{-1/2}$  parametrically, we find that the ``mass'' is
parametrically larger than $ {\lambda}^{-1/2}$, which in turn implies
that the function $f(r)$ that appears in the SdS metric changes sign.
At this point, the view of the universe as a cavity no longer applies.
\eqn{finequality}{ f(r) = 1 - \frac{2M}{r} - \frac{\lambda r^2}{3}< 0
\ .}
This implies an exchange of roles between the radial and time
coordinates in the description of the
space-time geometry.
The metric can then be rewritten by introducing $
T \equiv r$ or $T \equiv -r$ and $y\equiv t$
\eqn{newmetric}{ ds^2 = \frac{1}{g(T)}dT^2 - g(T)dy^2 - T^2
d\Omega_2^2\ ,}
where
\eqn{g(T)}{g(T) = \frac{\lambda T^2}{3} + \frac{2M}{T} - 1 \ .}
The space-time geometry is now time dependent. The inside
of the ``star'' and the outside geometry
either both collapse to a big crunch or expand from a big bang
singularity. The outside geometry looks like a
contracting or expanding hyper-cylinder. This is rather different from
the case where the entropy of the ``star''
satisfies the bound.

We do not expect that an increase of the entropy beyond the
bound, entails such drastic changes. Instead,
we believe that the excess entropy will be stored, some in small
black holes, but most we believe stored on the de
Sitter horizon.
The case concerning the behavior of a ``star'' has to
be contrasted with the case of black holes
where the entropy storage is maximal. A somewhat related phenomenon
is the case of the repulsion between two
black holes in de Sitter space \cite{usualsuspects}, with respective
areas  smaller than a  bound allowed by de
Sitter space, but whose merger would have created a black hole with
size larger than the bound

The main point that we are trying to argue in this paper is that
the entropy in radiation that a local
observer can measure cannot exceed a bound, that is smaller
parametrically than the Bekenstein bound. Any excess
entropy gets repackaged. So far we have provided circumstantial
evidence for the bound and what remains to be
shown is how this bound emerges for a local observer immersed in a
universe filled with
radiation and in the presence of a small cosmological constant. An
argument for the bound might be based on the
theorem \cite{bousso} that states that there is an upper bound on the
entropy and hence the ``mass'' of a black hole in de Sitter
space. Such a
black hole has a maximum ``size'' ${\lambda}^{-1/2}$. We have seen
earlier that in the absence of a
cosmological constant, the entropy that radiation in equilibrium can
fit into a sphere of size
${\lambda}^{-1/2}$, before the onset of black hole formation is of
$O({\lambda}^{-3/4})$ . Any more entropy
stored in radiation would need more space or be repackaged. The
cosmological constant merely acts as a cavity
which the radiation fills.

Before concluding it would be important to describe the observation
of a local observer immersed in a geometry determined by
a perfect fluid and a positive cosmological constant, without 
using a ``star'' to obtain the bound.
Although this gauge, known as the ``O-gauge'',
\cite{LennyO} can be constructed, the technical difficulty (for us) in
this gauge is that the dynamics is described by two dimensional
partial differential equations \footnote{see the appendix}. This is
in contrast with our use of a ``star'' where only one
dimensional differential equations are involved.

\section{\bf Conclusions}

Let us briefly recapitulate our main message: a local
observer in a four dimensional FRW universe with a positive
cosmological constant will never ``see''  more entropy in
the microwave background than $S \sim {\lambda}^{-3/4}$.
Any excess entropy up to $O({\lambda}^{-1})$ gets processed
into black holes or gets absorbed by the horizon. If we
ascribe the observed acceleration of our universe to a cosmological constant
then the entropy in the microwave background must be smaller than
$O(10^{91})$.

The local observer does not in general witness catastrophic events, he or she
merely sees the microwave background and the occasional black hole.
The experience of the observer becomes drastically different in the
collapsing phase if the amount of
entropy exceeds the ``Bekenstein bound'' of $O({\lambda}^{-1})$, he or
she ends with the rest of the universe in a big crunch.

The question of why the entropy in the microwave background is
$O(10^{85})$ which is some 
five orders of magnitude
smaller than the bound discussed in this paper remains unanswered.
A plausible answer is that if there is an excess
entropy produced, in reheating the universe, the process of repackaging
that excess entropy somehow overshoots the bound.
In this context it is interesting to note that the current estimate of
the entropy stored in black holes at the center of galaxies is
$O(10^{95})$.
Another possible explanation
in the context of inflation is that the subsequent reheating of the
universe  produces an amount of entropy short of the
bound.

\section{Acknowledgments}

   The research of Willy Fischler, Amit Loewy and Sonia Paban was
supported in part by NSF grant 0071512, and grant support from the US Navy, Office of Naval Research,
Grant No. N00014-03-1-0639, Quantum Optics Initiative.
Useful conversations
with Julie Blum, Eduardo di Napoli, Hyukjae Park and Marija Zanic are gratefully
acknowledged.

\section{Appendix}

In order to describe what a local observer experiences, it is useful
to work in the ``O-gauge'' coordinate system which only covers
the causal region that the observer. It was shown in
\cite{LennyO} that given a space-time metric of the FRW
form
\begin{equation} \label{FRWmetric}
ds^2 =  dt^2 -a(t)^2 \left( d \theta^2 + \sin^2 \theta
\ d\Omega_2^2 \right) \ ,
\end{equation}
where $-\infty < t < \infty$ and $0 \le \theta \le \pi$, one can make
the following coordinate transformation to a coordinate system that
covers only the causal patch with a metric of the form
\begin{equation}\label{Ogauge}
ds^2 =  e^v (dt^2 - d r^2) - Z^2 d\Omega_2^2 \ ,
\end{equation}
where $v$ and $Z$ are in general functions of space and time.

First, conformally rescale (\ref{FRWmetric}) by defining the new time
coordinate $\tau = \int dt/a(t)$.
The Penrose diagram associated with the new metric is depicted
below.
\begin{figure}[htbp]
\leavevmode
\centerline{
\psfig{figure=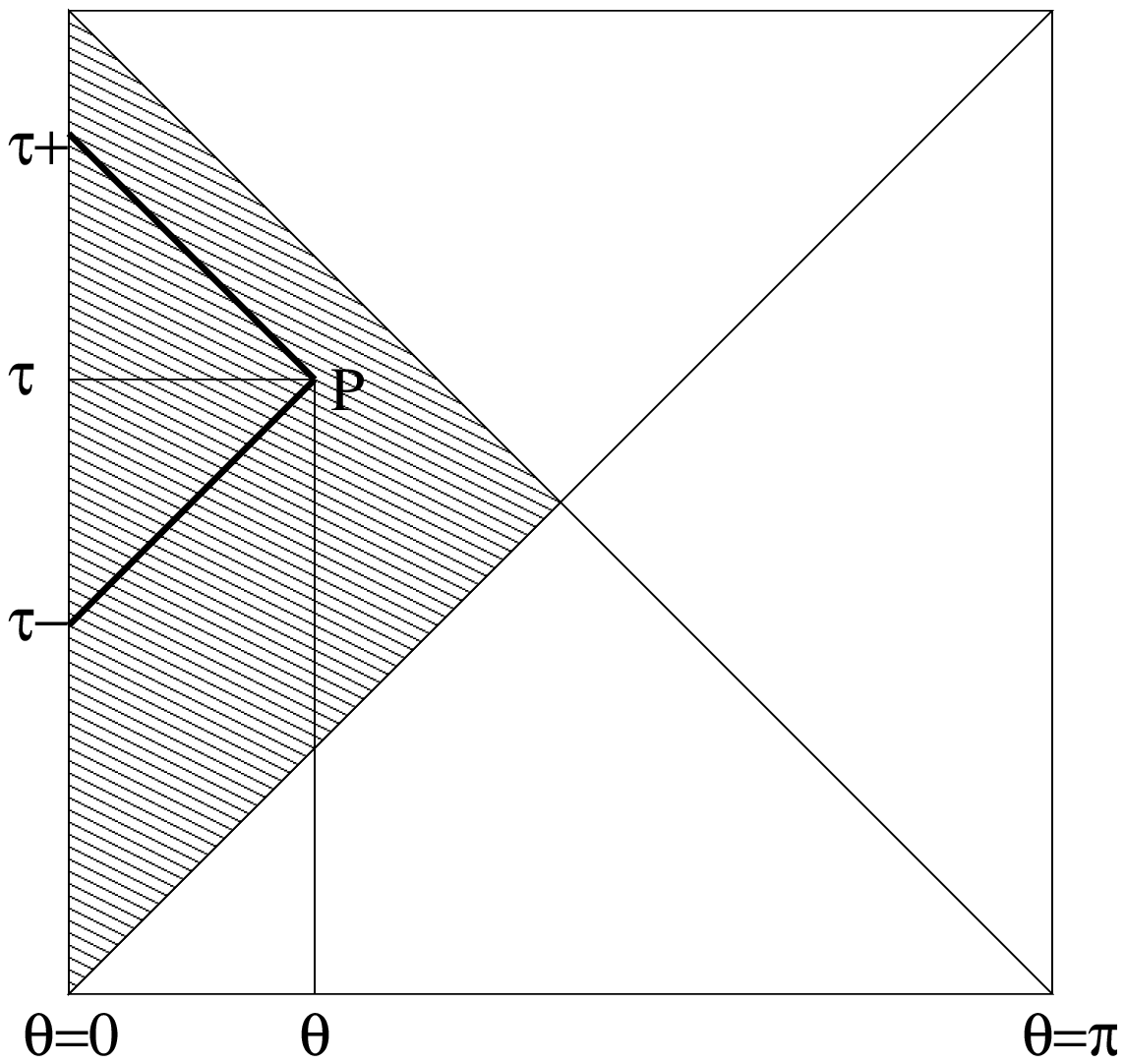,width=7cm,height=7cm,clip=}
}
\end{figure}
For a stationary observer at $\theta=0$, the causal patch is
marked by the shaded region. A point $P$ in the causal patch can be
mapped to a pair of points on the observer's world-line by following
the light-like trajectories to $\tau^+$ and $\tau^-$. Clearly, the
following relations hold $\tau = (\tau^+ + \tau^-)/2$ and
$\theta = (\tau^+ - \tau^-)/2$.
  From this it follows that the metric on the causal patch can
be written as
\begin{equation}
ds^2 = a^2 \left(\frac{\tau^+ + \tau^-}{2} \right) d\tau^+ d\tau^-
- a^2 \left(\frac{\tau^+ + \tau^-}{2} \right) \sin^2
\left(\frac{\tau^+ - \tau^-}{2} \right) d\Omega_2^2 \ ,
\end{equation}
where $\tau_{min} \le \tau^{\pm} \le \tau_{max}$.
Applying a further (holomorphic) coordinate transformation
$\tau^{\pm} \rightarrow X^{\pm}(\tau^{\pm})=t \pm r$, subject to the
gauge-fixing conditions $v(r=0)=Z(r=0)=0$
brings the metric to the form (\ref{Ogauge}).

We will now write Einstein's equations in the presence of a perfect
fluid with the
equation of state $p=\kappa \rho$ and a positive cosmological constant .
The energy-momentum tensor of this fluid is given by
\begin{equation}
T^{\mu \nu} = \rho \left( (1+\kappa) U^{\mu}U^{\nu}
+\kappa g^{\mu \nu} \right) \ ,
\end{equation}
where $U$ is the 4-velocity of the fluid. 
The $tt$, $rr$, and $tr$
components of Einstein's equation read
$$
\frac{\Zdot \vdot}{Z} +\frac{\Zdot^2}{Z^2} +
\frac{v'Z'}{Z}-2\frac{Z''}{Z}+\frac{e^v}{Z^2}-\frac{Z'^2}{Z^2}
-\lambda e^v= T_{tt} \ ,
$$
$$
\frac{\Zdot \vdot}{Z}-\frac{\Zdot^2}{Z^2}-2\frac{\ddot{Z}}{Z}
+\frac{v'Z'}{Z}-\frac{e^v}{Z^2}+\frac{Z'^2}{Z^2}
+\lambda e^v = T_{rr} \ ,
$$
\begin{equation}
\frac{\Zdot v'}{Z} + \frac{Z' \vdot}{Z} -2 \frac{\Zdot'}{Z}
= T_{tr} \ .
\end{equation}
The last two components, $\theta \theta$ and $\phi \phi$ give the same
equation (assuming spherical symmetry)
\begin{equation}
e^{-v} \left( Z \ddot{Z} - Z Z'' + \frac{1}{2} Z^2 \ddot{v} -
\frac{1}{2} Z^2 v'' - e^v \right) + \lambda Z^2 
= T_{\theta \theta} \ .
\end{equation}
The system is further constrained by energy-momentum conservation
$\nabla_{\mu}T^{\mu \nu}=0$.
One must now give initial conditions
on a space-like hyper-surface $t=t_0$ for the geometry and the
fluid.
The entropy on such an initial slice is given by
\begin{equation}
S(t=t_0) \sim \int dr \ e^{v/2}Z^2 \ \rho^{\frac{1}{1+\kappa}} \ .
\end{equation}
What, in this gauge, would be
the signal that the bound has been violated? \cite{FLPprogress}.

%


\end{document}